\documentclass[prx,aps,amssymb,twocolumn, superscriptaddress]{revtex4-2}
\usepackage{amsmath}
\usepackage{amssymb}
\usepackage{amsthm}
\usepackage{amsfonts}
\usepackage{listings}
\usepackage{physics}
\usepackage{enumerate}
\usepackage{latexsym}
\usepackage{color}
\usepackage{bm}
\usepackage{hyperref}
\hypersetup{
 pdfnewwindow=true, colorlinks=true,
 linkcolor=blue, anchorcolor=blue,
 citecolor=blue, filecolor=blue,
 menucolor=blue, urlcolor=blue}

\usepackage{psfrag}

\usepackage{bm}
\usepackage{graphicx}
\usepackage{subfigure}

\RequirePackage[normalem]{ulem} 
\RequirePackage{color}\definecolor{RED}{rgb}{1,0,0}\definecolor{BLUE}{rgb}{0,0,1}

\newcommand{\bk}{{\bf k}}
\newcommand{\be}{{\vb* e}}
\newcommand{\bs}{{\vb* s}}
\def\eg{{\it e.g.}\ }

\input{epsf}

\newcommand{\nc}{\newcommand}
\nc{\webirvsp}{\href{https://github.com/zjwang11/irvsp}{\texttt{IRVSP}} }
\nc{\webirtb}{\href{https://github.com/zjwang11/irvsp}{\texttt{ir2tb}} }
\nc{\webirpw}{\href{https://github.com/zjwang11/ir2pw}{\texttt{ir2ph}} }
\nc{\webchecktopmat}{\href{https://www.cryst.ehu.es/cryst/checktopologicalmagmat}{\texttt{Check Topological Mat}}}
\nc{\webposabr}{\href{https://github.com/zjwang11/UnconvMat/blob/master/src_pos2aBR.tar.gz}{\texttt{POS2ABR}} }
\nc{\webUnconvMat}{\href{http://tm.iphy.ac.cn/UnconvMat.html}{\texttt{UnconvMat}} }
\nc{\online}{\href{http://tm.iphy.ac.cn/UnconvMat.html}{online}}

\begin{document}

\tolerance 10000

\newcommand{\vk}{{\bf k}}

\draft

\title{Majorana corner modes in unconventional monolayers of the 1$T$-PtSe$_2$ family}
%\title{Majorana corner modes and symmetry indicator-free unconventionaliy in 1T-monolayers of PtSe$_2$ family}

%in English titles articles and words like to, on, at etc are always spelled with small letters
\author{Haohao Sheng}
%\thanks{These authors contributed equally to this work.}
\affiliation{Beijing National Laboratory for Condensed Matter Physics,
and Institute of Physics, Chinese Academy of Sciences, Beijing 100190, China}
\affiliation{University of Chinese Academy of Sciences, Beijing 100049, China}

\author{Yue Xie}
%\thanks{These authors contributed equally to this work.}
\affiliation{Beijing National Laboratory for Condensed Matter Physics,
and Institute of Physics, Chinese Academy of Sciences, Beijing 100190, China}
\affiliation{University of Chinese Academy of Sciences, Beijing 100049, China}

\author{Quansheng Wu}
\affiliation{Beijing National Laboratory for Condensed Matter Physics,
and Institute of Physics, Chinese Academy of Sciences, Beijing 100190, China}
\affiliation{University of Chinese Academy of Sciences, Beijing 100049, China}

\author{Hongming Weng}
\affiliation{Beijing National Laboratory for Condensed Matter Physics,
and Institute of Physics, Chinese Academy of Sciences, Beijing 100190, China}
\affiliation{University of Chinese Academy of Sciences, Beijing 100049, China}

\author{Xi Dai}
\affiliation{Department of Physics, Hong Kong University of Science and Technology, Hong Kong 999077, China}

\author{B. Andrei Bernevig}
\affiliation{Department of Physics, Princeton University, Princeton, New Jersey 08544, USA}

\author{Zhong Fang}
\affiliation{Beijing National Laboratory for Condensed Matter Physics,
and Institute of Physics, Chinese Academy of Sciences, Beijing 100190, China}
\affiliation{University of Chinese Academy of Sciences, Beijing 100049, China}

\author{Zhijun Wang}
\email{wzj@iphy.ac.cn}
\affiliation{Beijing National Laboratory for Condensed Matter Physics,
and Institute of Physics, Chinese Academy of Sciences, Beijing 100190, China}
\affiliation{University of Chinese Academy of Sciences, Beijing 100049, China}

\begin{abstract}
In this work, we propose that Majorana zero modes can be realized at the corners of the two-dimensional unconventional insulator. We demonstrate that 1$T$-PtSe$_2$ is a symmetry indicator-free (SI-free) unconventional insulator, originating from orbital hybridization between Pt $d$ and Se $p_{x,y}$ states. The kind of  SI-free unconventionality has no symmetry eigenvalue indication. Instead, it is diagnosed directly by the Wannier charge centers by using the one-dimensional Wilson loop method. 
The obstructed edge states exhibit strong anisotropy and large Rashba splitting. By introducing superconducting proximity and an external magnetic field, the Majorana corner modes can be obtained in the 1$T$-PtSe$_2$ monolayer. In the end, we construct a two-Bernevig-Hughes-Zhang model with anisotropy to capture the Majorana physics.
\end{abstract}

\maketitle

\section{Introduction}
%\paragraph*{Introduction.}

\begin{figure}[!t]
\centering
\includegraphics[width=6 cm]{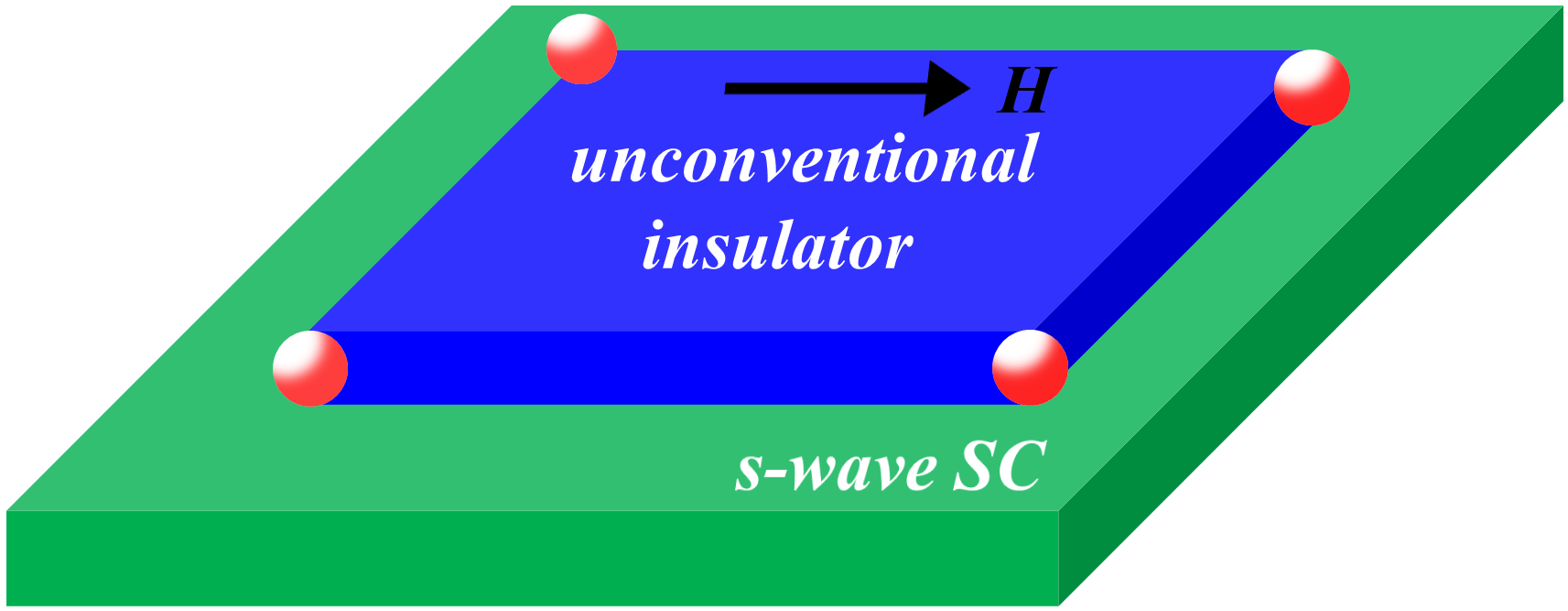}
\caption{(Color online) The schematic of a heterostructure composed of an unconventional insulator/OAI with Rashba edge states on top of an $s$-wave superconductor. Majorana corner modes can be induced under an external magnetic field.
} \label{fig-MZM1}
\end{figure}

Majorana zero modes (MZMs) in topological superconductors \cite{Majorana} have attracted great interest in the past two decades and an amount of physical systems \cite{Kitaev,TSC1,TSC2,TSC3,TSC4,TSC5} has been proposed to realize Majorana modes. Among them, the Rashba semiconducting nanowire, which hosts Majorana end states with superconducting proximity and under a magnetic field, is a well-studied system \cite{1Dwire,nanowire}. Recently, the idea of higher-order topological superconductors \cite{SOTSC,xie2023hybrid,type2-PhysRevB.105.L041105,huang2024surface} broadens the research area of Majoranas and turns the Majorana modes to corners and hinges. Normally, higher-order topological superconductivity always requires either unconventional pairing order \cite{pwave1,pwave2,dwave,sextendwave,withmag,TIsextend,QSHIsextend} ($p$-, $d$- or $s_\pm$-wave) or complicated junctions \cite{junction}, which are difficult to implement in experiment. By using topologically protected edge states of quantum spin Hall insulators in proximity contact with an $s$-wave superconductor, Majorana corner modes can be realized when subjected to an external magnetic field or ordering \cite{PRL.124.227001,QSHImag2}. However, the obstructed edge states of unconventional insulators (obstructed atomic insulators; OAI)~\cite{aBR2021,aBR2022,real_space_invariants2021,tqc2017,PhysRevLett.121.126402,song2020rsi,xu2021filling,Li2111, ptte1.75, zhang2022large, xu2022catalogue, zhang2023ph, Nature.Josephson, YangziLong2023}, viewed as a one-dimensional (1D) system exhibiting strong Rashba band splitting, also have potential to host Majorana corner modes on superconducting and magnetic substrates, which is an unexplored area (Fig.~\ref{fig-MZM1}).

As a representative of 1$T$-phase transition metal dichalcogenides ($MX_2;~M=$ Ni, Pd, Pt; $X=$ S, Se, Te), 1$T$-PtSe$_2$ shows many excellent characteristics, such as high electron mobility~\cite{ptse-mobility}, helical spin texture~\cite{ptse-spin}, unique magnetic ordering~\cite{ptse-mag}, and excellent photocatalytic activity~\cite{ptse-cata}, etc~\cite{ptse-Dirac2,ptse-pat,tyner2023solitons}. Recently, the edge electronic states of PtSe$_2$ zigzag ribbon have been predicted theoretically~\cite{ptse-edge1} and subsequently confirmed experimentally~\cite{ptse-edge2}. The PtSe$_2$ ribbon is demonstrated to grow naturally along the zigzag direction, and the edge states make few-layer PtSe$_2$ more semimetallic~\cite{ptse-edge2,PtSe2023edge}. However, the formation mechanism of the edge states is still unrevealed.

In this work we first propose that MZMs can be realized at the corners of the unconventional 1$T$-PtSe$_2$ monolayer. Based on the first-principles calculations, we demonstrate that 1$T$-PtSe$_2$ is a symmetry indicator-free (SI-free) unconventional insulator. This kind of unconventionality is not indicated by any symmetry eigenvalues, but is directly diagnosed by the computed Wannier charge centers (WCCs). 
Our detailed analysis shows that the unconventional nature is attributed to the splitting of the $E@2d$ band representation (BR) of Se $p_{x,y}$ states. 
The obstructed edge states exhibit strong anisotropy and large Rashba splitting. By introducing superconducting proximity and an external magnetic field, the MZMs can be generated at the corners of 1$T$-PtSe$_2$ monolayer.
The magnetic substrate CrGeTe$_3$ is proposed to induce a sizable gap in the Dirac edge states.

\section{Methodology}
We carried out the first-principles calculations based on the density functional theory (DFT) with projector augmented wave (PAW) method~\cite{paw1, paw2}, as implemented in the Vienna \emph{ab initio} simulation package (VASP)~\cite{KRESSE199615, vasp}. The generalized gradient approximation (GGA) in the form of Perdew-Burke-Ernzerhof (PBE) function~\cite{pbe} was employed for the exchange-correlation potential. The kinetic energy cutoff for the plane wave expansion was set to 450 eV. The Brillouin zone was sampled by Monkhorst-Pack method in the self-consistent process, with an 18$\times$18$\times$10 $k$-mesh for PtSe$_2$ bulk and a 10$\times$1$\times$1 (1$\times$6$\times$1) $k$-mesh for PtSe$_2$ ribbon with zigzag (armchair) edge.
The thickness of the vacuum was set to \textgreater ~20 \AA ~for PtSe$_2$ ribbon with a zigzag (armchair) edge. The irreducible representations (irreps) of electronic states were obtained from the program \webirvsp~\cite{irvsp}. With the obtained \texttt{tqc.data}, elementary band representations/atomic valence-electron band representations decomposition of the band structure was done on the \webUnconvMat website~\cite{aBR2022}. The maximally localized Wannier functions for the Pt $d$ and Se $p$ orbitals were constructed using the WANNIER90 package~\cite{wannier90}.

\section{Calculational Results}

\begin{table}[!b]
    \centering
    \caption{The atomic valence-electron band representations (ABRs) of 1$T$-PtSe$_2$. The Pt $d$ orbitals form the $A_{1g}$, $E_g$, and $E_g$ irreps at the Pt ($1a$) site, being $A_{1g}@1a$, $E_{g}@1a$, and $E_{g}@1a$ ABRs; while the Se $p$ orbitals form the $A_1$ and $E$ irreps at the Se ($2d$) site, being $A_1@2d$ and $E@2d$ ABRs.}
    \label{table:aBRs}
    \begin{ruledtabular}
    \begin{tabular}{cccrcc}
    %\hline\hline
        Atom & WKS($q$) & Symm. &Conf.& Irreps($\rho$) & ABRs($\rho@q$) \\
        \hline
        Pt & $1a$ & -3m & $d^{10}$& $d_{z^2}$:$A_{1g}$ & $A_{1g}@1a$\\
        &(000) & & &$d_{xy,x^2-y^2}$:$E_g$ & $E_g@1a$\\
        & & & &$d_{xz,yz}$:$E_g$ & $E_g@1a$\\
        \hline
        Se & $2d$ & 3m  &  $p^4$&$p_z$:$A_1$ & $A_1@2d$\\
        & $(\frac{1}{3}\frac{2}{3}z)(\frac{2}{3}\frac{1}{3}$-$z)$ & && $p_{x,y}$:$E$ & $E@2d$\\
        \end{tabular}
        \end{ruledtabular}
\end{table}

\subsection{Electronic band structure and band representation}
\begin{figure}[!t]
\centering
\includegraphics[width=8.5 cm]{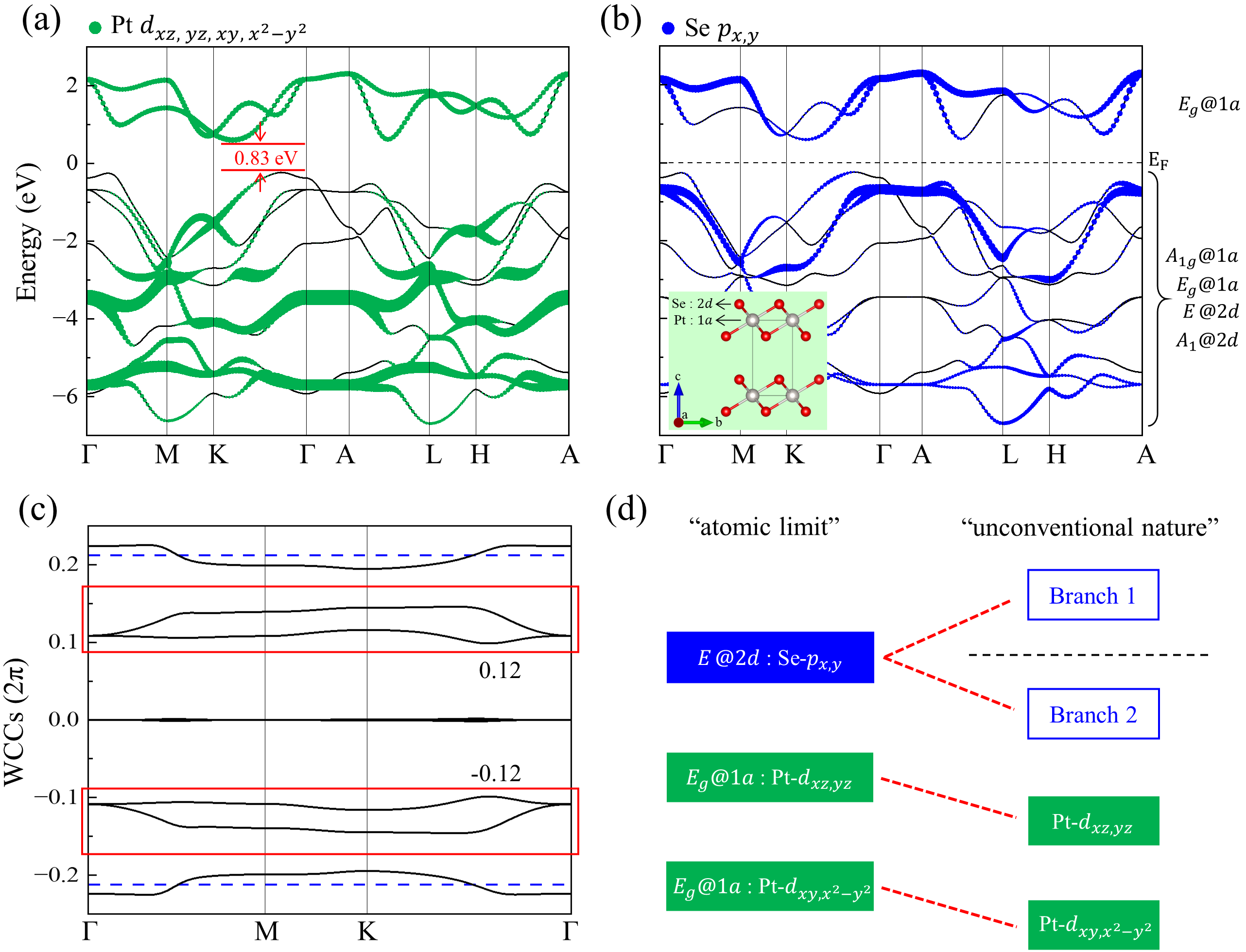}
\caption{(Color online)
(a,b) Orbital-resolved band structures of 1$T$-PtSe$_2$. The sizes of (a) green and (b) blue circles represent the weights of Pt $d_{xz,yz,xy,x^2-y^2}$ and Se $p_{x,y}$ orbitals, respectively. ABR decomposition is shown in (b). The inset of (b) presents the side view of 1$T$-PtSe$_2$.
(c) The $z$-directed Wannier charge centers (WCCs) for the occupied nine bands. The blue dashed lines ($z=\pm 0.21c$) indicate the locations of the Se atoms.
(d) The schematic of band hybridization, giving rise to the symmetry indicator-free (SI-free) unconventional nature. At the atomic limit (without interatom hybridization), all the bands originate from ABRs. After band hybridization in crystals, the occupied bands may form the BRs that are not located at the atoms, which is termed unconventionality. 
SI-free unconventionality cannot be diagnosed by BR analysis or symmetry indicator. The colors of the filled boxes indicate the atomic orbital characters, while the unfilled boxes represent the nonatomic states.
} \label{fig-PtSebulk}
\end{figure}

%\paragraph*{Electronic band structure and band representation analysis.}
The 1$T$-phase PtSe$_2$ possesses a structure with space group $P\overline{3}m1$, where each Pt atom lies at the center of an octahedral cage formed by Se atoms, as shown in inset of Fig.~\ref{fig-PtSebulk}(b). The layers are connected by van der Waals force. The Pt and Se atoms are located at the $1a$ and $2d$ Wyckoff positions, respectively. 
In topological quantum chemistry theory~\cite{tqc2017}, the BR $\rho@q$ is induced by the $\rho$-irrep orbital at the $q$ site in lattice space, which also indicates a set of $k$-irreps in momentum space. The generators of the BRs are elementary band representations (EBRs).  
Atomic valence-electron band representations (ABRs) are defined as the BRs induced by the atomic valence electrons~\cite{aBR2022}. In PtSe$_2$, all the low energy bands originate from the valence electrons of Pt $d$ and Se $p$ orbitals, which form the ABRs of the compound.
These ABRs are generated by \webposabr and presented in TABLE~\ref{table:aBRs}. 
The band structure of PtSe$_2$ is obtained with an indirect gap of 0.83 eV in Fig.~\ref{fig-PtSebulk}(a). Based on symmetry eigenvalues (or irreps), the nine valence bands can be decomposed to the sum of ABRs: $(A_1+ E)@2d + (E_g +A_{1g})@1a$, while the two conduction bands belong to $E_g@1a$~\cite{tqc2017,ashvin-nc2017-symm}. 
They are exactly the obtained ABRs of TABLE~\ref{table:aBRs}.
The ABRs decomposition seems consistent with the valence states of Pt$^{4+}$ and Se$^{2-}$, implying that the Se $p$ orbitals are fully occupied and two conduction bands are mainly from the Pt $d$ orbitals of the $E_g$ irrep.

\begin{table}[!b]
\begin{ruledtabular}
\caption{The $k$-irreps of the ABR $E@2d$ are presented below. They can be decomposed into two parts.}
   \begin{tabular}{c c c }%
          $E@2d$      & Branch 1        & Branch 2       \\
          &(identical to $E_g@1a$)& \\
            \hline
            $\Gamma$$_3^+ \oplus \Gamma$$_3^-$     &  $\Gamma$$_3^+$        & $\Gamma$$_3^-$                  \\
            A$_3^+ \oplus$ A$_3^-$                 &  A$_3^+$               & A$_3^-$                           \\
            H$_1 \oplus$ H$_2 \oplus$ H$_3$        &  H$_3$                 & H$_1 \oplus$ H$_2$                \\
            K$_1 \oplus$ K$_2 \oplus$ K$_3$        &  K$_3$                 & K$_1 \oplus$ K$_2$              \\
            L$_1^+ \oplus$ L$_1^- \oplus$ L$_2^+ \oplus$ L$_2^-$       & L$_1^+ \oplus$ L$_2^+$    & L$_1^- \oplus$ L$_2^-$   \\
            M$_1^+ \oplus$ M$_1^- \oplus$ M$_2^+ \oplus$ M$_2^-$       & M$_1^+ \oplus$ M$_2^+$    & M$_1^- \oplus$ M$_2^-$    \\
  \end{tabular}
\label{table:2}
\end{ruledtabular}
\end{table}

\subsection{SI-free unconventional nature}
On the other hand, we calculate the orbital-resolved band structures in Figs.~\ref{fig-PtSebulk}(a) and \ref{fig-PtSebulk}(b). 
The Pt $d_{xz,yz,xy,x^2-y^2}$ orbitals form two $E_g$ irreps at the $1a$ site (being $E_g@1a$ ABR), while the Se $p_{x,y}$ orbitals form an $E$ irrep at the $2d$ site (being $E@2d$ ABR).
Figure ~\ref{fig-PtSebulk}(a) shows most weights of the $E_g$-irrep Pt $d$ orbitals below the Fermi level ($E_F$).  In contrast, many weights of the $E$-irrep Se $p$ orbitals are above the $E_F$ in Fig.~\ref{fig-PtSebulk}(b). These results are not consistent with Pt$^{4+}$ and Se$^{2-}$ states at all. 

%To be more precise
By using the 1D Wilson loop method~\cite{Ben-nc2020-wcc, Ben2023spinresolved}, the $z$-directed WCCs are computed, indicating the electronic locations along the $z$ direction directly. The $(A_{1g}+E_g)@1a$ bands should have three WCCs at $z=0c$, while the $(E+A_1)@2d$ bands should have WCCs at $z=\pm 0.21c$ [dashed lines in Fig.~\ref{fig-PtSebulk}(c)], being aligned with Se positions. The WCCs for nine occupied bands are obtained in Fig.~\ref{fig-PtSebulk}(c), and the results show that the average of the two Wilson bands in the red box is $z= 0.12c$, quite far away from the Se atoms. This indicates the unconventional nature of mismatch between the Wannier/electronic charge centers and atomic positions in PtSe$_2$.

\subsection{Origin of unconventional nature} 
To investigate the origin of unconventional nature, we have checked the irreps carefully, because only the bands of the same irreps can hybridize.
The schematic of band hybridization in PtSe$_2$ is given in Fig.~\ref{fig-PtSebulk}(d). 
Starting from the atomic limit, the energy levels of $d$ orbitals ($E_g@1a$) are lower than that of $p_{x,y}$ orbitals ($E@2d$) in the Wannier-based tight-binding Hamiltonian extracted from the DFT calculations, giving rise to a half filling of $E@2d$ ABR at $E_F$.
Then, the $E@2d$ ABR can be decomposed into two separate branches with branch 1 identical to $E_g@1a$, as presented in TABLE~\ref{table:2}.
Due to the hybridization between Pt $d$ ($E_g@1a$)and Se $p_{x,y}$ (branch1 of $E@2d$), branch 1 is fully unoccupied and branch 2 is fully occupied. 
The hybridization process gives rise to the SI-free unconventionality characterized by the offset charge centers, which is consistent with the obtained WCCs in the 1D Wilson loop method.

This SI-free unconventionality is not protected by any symmetry, being of accidental obstructed atomic limit. Although it seems that the SI-free unconventionality can be removed by changing the relative onsite energy without closing the energy gap, the substitution with the same group elements usually does not qualitatively change the relative onsite energy at all.
As a matter of fact, all the 1$T$-$MX_2$ compounds of the PtSe$_2$ family  possess the SI-free unconventionality. First, the onsite energy of $M$ $d$ orbitals is lower than that of $X$ $p_{x,y}$ orbitals for all 1$T$-$MX_2$ members at the atomic limit. Second, the interlayer hybridization of an $X$ $p_z$ orbital may lead to metallicity in some 1$T$-$MX_2$ compounds~\cite{shengMTe2}, which does not really affect the SI-free unconventionality (relying on the $X$ $p_{x,y}$ orbitals). As a result, the obstructed edge states are always expected in the PtSe$_2$-family compounds.
Recently, the existence of SI-free unconventionality and obstructed hinge states is demonstrated in NiTe$_2$ bulk, and the magnetic field filtering of hinge supercurrent in NiTe$_2$-based Josephson junctions is observed experimentally~\cite{nite2}.

\begin{figure}[!t]
\centering
\includegraphics[width=8.5 cm]{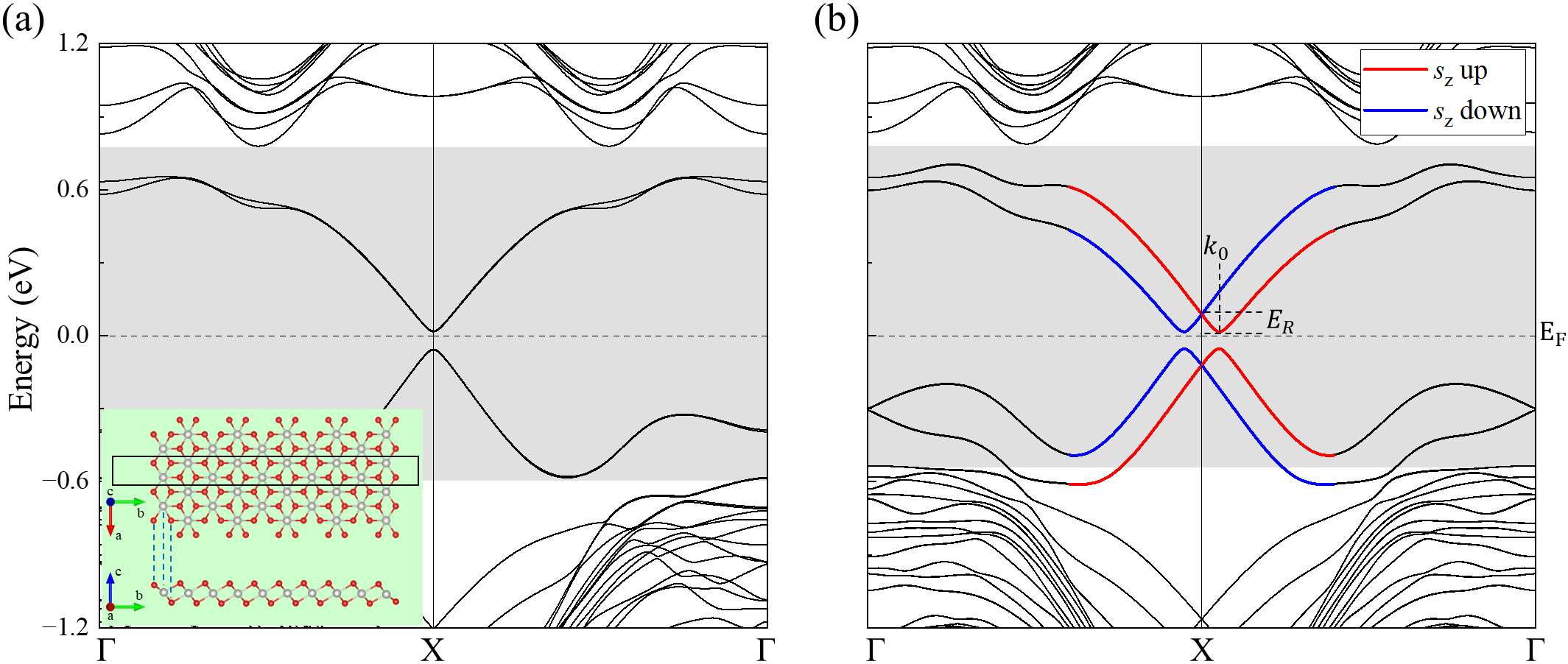}
\caption{(Color online)
Band structures (a) without and (b) with SOC of PtSe$_2$ ribbon with zigzag edge. Inset of (a) presents the structure of the zigzag edge. The black rectangle represents the unit cell. Red and blue bands in (b) indicate the $s_z$ up and $s_z$ down channels due to SOC. The $k_0$ and $E_R$ denote the momentum offset and Rashba energy, respectively. 
} \label{fig-zigzag}
\end{figure}

\subsection{Obstructed edge states}
Due to the unconventional nature of the electronic band structure, obstructed edge states can emerge. 
The band structure of the zigzag edge ($x$ direction) is presented in Fig.~\ref{fig-zigzag}. Without spin-orbit coupling (SOC), the edge states appear in the bulk band gap (shadowed area), as shown in Fig.~\ref{fig-zigzag}(a). The obstructed edge states behave as a massive Dirac band around the $X$ point, rather than a parabolic band, because the linear dispersion survives in a wide range ($\sim \frac{1}{3}$) of the edge Brillouin zone. The gap of the edge Dirac band is 0.07 eV. The edge gap is consistent with the previous theoretical and experimental results~\cite{ptse-edge1,ptse-edge2,PtSe2023edge}. In addition, the band structure of the armchair edge ($y$ direction) is also depicted in Fig.~\ref{figS-zig+arm} of Appendix~\ref{SM1}. Instead of a massive Dirac band, the obstructed edge states show an M-shaped band. Moreover, the armchair edge gap is larger than the zigzag edge gap, exhibiting strong anisotropy.

Upon including SOC, the zigzag edge states will split into two massive Dirac bands, exhibiting large Rashba splitting in Fig.~\ref{fig-zigzag}(b). The coupling strength of the Rashba SOC can be obtained as $\alpha_{R}=\frac{2E_{R}}{k_0}=3.35$ eV \AA, where $k_0=0.043$ \AA  ~and $E_R=0.072 $eV denote the momentum offset and Rashba energy, respectively. The remarkable $\alpha_{R}$ is as large as those in the  BiTeI~\cite{BiTeI} and Bi/Ag(111) surface alloy ~\cite{Bi-Ag-PRL}.
As a result, two massive Dirac bands are supposed to have the opposite spin. Due to the coexistence of time reversal and $M_x$ symmetries, $\ev{s_x}=0$ is required. The $x$-directed magnetic field can easily lift the Kramers' degeneracy at the $X$ point. Additionally, $\ev{s_y}$ and $\ev{s_z}$ are computed, as shown in Fig.~\ref{figS-spin} of Appendix~\ref{SM1}, which indicate that the $s_z$ component is dominant. 
We conclude that the obstructed edge states exhibit large Rashba splitting and strong anisotropy. 
The edge atoms of ribbons are fully relaxed in the calculations. These results are obtained by the \textit{ab initio} self-consistent calculations.

\subsection{Proposal for MZMs in the corners}
Due to the existence of obstructed edge states with large Rashba splitting and strong anisotropy, we propose that the MZMs can be realized at the corners of 1$T$-PtSe$_2$ monolayer. Under $s$-wave superconducting proximity and an $x$-directed magnetic field, the zigzag edge can be tuned to an equivalent spinless $p$-wave Kitaev chain~\cite{Kitaev}, while the armchair edge is fully gapped (Appendix~\ref{SM1}). 
Thus, at the corners (ends of the zigzag edges), MZMs can be realized with a proper chemical potential. 
It is proposed here that MZMs exist at the corners of the two-dimensional (2D) topologically trivial insulator with unconventionality.

\begin{figure}[!t]
\centering
\includegraphics[width=8.5 cm]{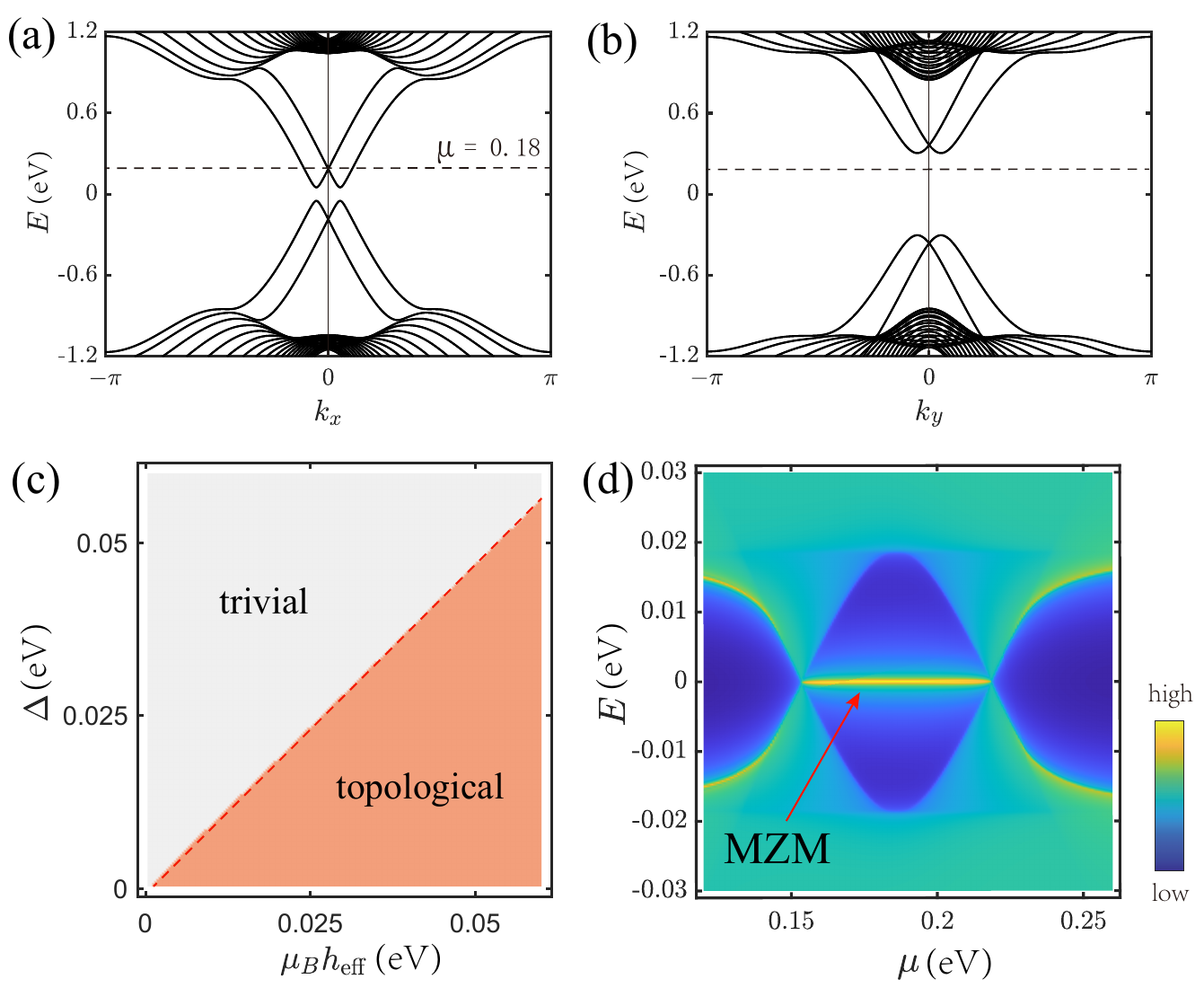}
\caption{(Color online)
Results of the two-BHZ model. (a) Energy dispersions in a ribbon geometry along $x$ ($L_y=20$ lattice sites). (b) Energy dispersions in a ribbon geometry along $y$ ($L_x=20$ lattice sites). (c) Phase diagram with superconducting pairing ($\Delta$) vs magnetic field ($h_{\mathrm{eff}}$) with chemical potential $\mu=0.18$ eV. The orange regime (topological) holds Majorana corner states, while the grey regime (trivial) does not. In (d), with $\mu_B h_{\mathrm{eff}}=0.04$ eV and $\Delta=0.02$ eV, we plot the corner spectrum evolution as a function of $\mu$. One can see that when tuning $\mu$ in the magnetic gap (0.15 eV$<\mu<$0.22 eV), the MZM is obtained at the corner.
} \label{fig-MZM}
\end{figure}

To simulate the Majorana corner modes in the system,  we construct a 2D lattice model, consisting of two copies of Bernevig-Hughes-Zhang (BHZ) Hamiltonians,
\begin{equation}\begin{aligned}
    H(\bk) = &\{m+b(\cos k_x+\cos k_y)\}\Gamma_{0z0} \\ &- \beta\sin k_x(\cos\theta\Gamma_{zxy} + \sin\theta\Gamma_{zxz}) \\ &+ \beta\sin k_y\Gamma_{0y0} + \xi\Gamma_{z00} - \alpha\Gamma_{xx0} + d\Gamma_{x00}.
\end{aligned}\end{equation}
Here, $\Gamma_{ijk}=\rho_i\sigma_js_k$ ($i,j,k=0,x,y,z$). Pauli matrices $\rho_i$,  $\sigma_j$ and $s_k$ act on BHZ,  orbital, and spin spaces, respectively. 
$\xi$ adds energy difference between the two copies of BHZ states. $d$ generates the anisotropy on two different edges.  $\alpha$ introduces the hybridization between the two copies of edge Dirac states. 
These parameters are estimated by fitting the band structure obtained from the first-principles calculations.
In Figs.~\ref{fig-MZM} (a) and \ref{fig-MZM} (b), we plot the $x$-directed (ZZ) and $y$-directed (AC) edge dispersions. 
In the two-BHZ model, when tuning the chemical potential $\mu$ to the degenerate states on ZZ edge, there are no Fermi-level states on the AC edge [$\mu=0.18$ eV; the dashed line in Figs.~\ref{fig-MZM} (a) and \ref{fig-MZM} (b)]. Therefore, only the ZZ edge responds to the external magnetic field $H_Z=\mu_Bh_{\text{eff}}\hat{\be}\cdot\bs$ ($\hat{\be}$ is unit vector and is chosen to be $\hat{\be}=e_x$ in the following) and superconducting proximity. The full Bogoliubov-de Gennes (BdG) Hamiltonian with magnetism is,
\begin{equation}\begin{aligned}
    H_{BdG}(\bk)=\left(\begin{array}{cc} H(\bk)+H_Z & -i\Delta s_y \\ i\Delta^*s_y & -H^*(-\bk)-H^*_Z \end{array}\right).
\end{aligned}\end{equation}
Both the Zeeman coupling $h_{\text{eff}}$ and superconducting pairing order $\Delta$ can open an energy gap at $\mu$. Specifically, they are competing with each other, giving rise to the phase diagram in Fig.~\ref{fig-MZM}(c).
When the Zeeman gap is dominant at $k_x=0$, the ZZ edge becomes an equivalent spinless $p$-wave Kitaev system.
In the topological region, we demonstrate that superconducting proximity can induce MZMs as shown in Fig.~\ref{fig-MZM}(d), which are bounded to the corners. 
Very recently, based on the obstructed surface states with huge Rashba SOC in unconventional insulator Nb$_3$Br$_8$~\cite{aBR2022,Nature.Josephson}, 2D topological superconductivity and Majorana states are proposed to be realized in the $s$-wave superconductor NbSe$_2$/Nb$_3$Br$_8$/ferromagnetic insulator CrI$_3$ heterostructure~\cite{hjn-PhysRevLett.132.036601}, which is the three-dimensional version similar to our proposal.

\subsection{Magnetic substrate CrGeTe$_3$}

\begin{figure}[!htb]
\centering
\includegraphics[width=8.5 cm]{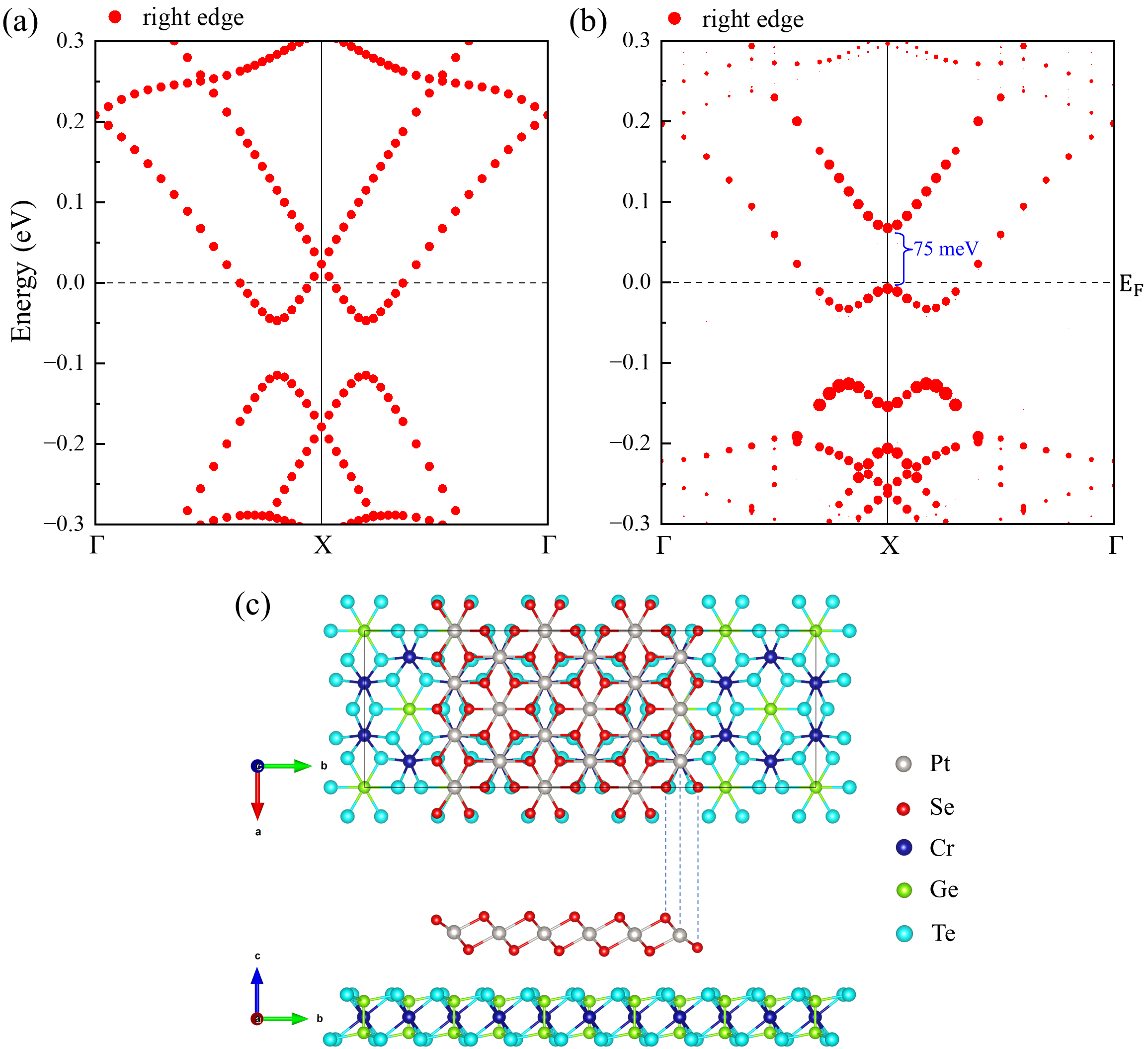}
\caption{(Color online)
Band structures with SOC of the PtSe$_2$ ribbon with zigzag edge on CrGeTe$_3$ substrate, hosting (a) an infinite interlayer distance and (b) an actual interlayer distance obtained by relaxation. The magnetic moment of CrGeTe$_3$ substrate is set parallel to the $x$ direction (along the zigzag edge direction). The size of red circle represents the weight of right edge in the PtSe$_2$ ribbon. The gap opened by Zeeman splitting is labeled. 
(c) Top and side views of the PtSe$_2$ ribbon with zigzag edge on CrGeTe$_3$ substrate. The black rectangle represents the unit cell. The right edge of the PtSe$_2$ ribbon is marked by dotted lines.
} \label{fig-PtSe+CrGeTe}
\end{figure}

To realize the Majorana corner modes in 1$T$-PtSe$_2$ monolayer, the effective Zeeman splitting of the zigzag edge states is important; in other words, the gap due to the external magnetic field at the X point matters in this proposal. Here, we propose that the ferromagnetic insulator CrGeTe$_3$~\cite{CrGeTe-nature,CrGeTe+PtSe2-PRB} is a proper substrate to induce a magnetic gap on the zigzag edge of 1$T$-PtSe$_2$ monolayer. 
The structure for this simulation is shown in Fig.~\ref{fig-PtSe+CrGeTe}(c). 
The results of Figs.~\ref{fig-PtSe+CrGeTe}(a) and \ref{fig-PtSe+CrGeTe}(b) show that the $x$-directed magnetic moment of CrGeTe$_3$ substrate can cause a large Zeeman gap of 75 meV at the X point. %(see Fig.~\ref{figS5} for $y$ and $z$-directed magnetic moments)
Covered by a conventional $s$-wave Bardeen-Cooper-Schrieffer superconductor, \eg aluminum, the MZMs at the corners of 1$T$-PtSe$_2$ monolayer are experimentally accessible.

\section{Conclusion}

The SI-free unconventionality of charge mismatch proposed here does not have any symmetry eigenvalue indication, analog to SI-free topological insulators~\cite{ashvin-nc2017-symm}. 
But it can be directly diagnosed by the computed WCCs using the 1D Wilson loop method. 
We demonstrate that the SI-free unconventionality is attributed to the splitting of the decomposable BR with band hybridization.  The SI-free unconventionality widely exists in the compounds with (Ga/In)$^{3+}$ or (Ge/Sn)$^{4+}$ valence states, which suggests that the $s$-orbital states would be empty and all above the $E_F$. On the contrary, the DFT results show that these states are quite far below the $E_F$. The detailed BR analyses and orbital-resolved band structures of SnSe$_2$, CuInSe$_2$, and Cu$_2$ZnGeSe$_4$ are presented in Appendix~\ref{SM2}. These SI-free unconventional compounds have exhibited superconductivity~\cite{snSC1,snSC2}, and optoelectronic or solar cell applications~\cite{CuInSe2-solarcell,Cu2ZnGeSe4-solarcell}.

We demonstrate that 1$T$-PtSe$_2$ is a SI-free unconventional insulator, whose unconventional nature originates from the band hybridization between Pt $d$ and Se $p_{x,y}$ states.
The obstructed electronic states on the zigzag edge exhibit large Rashba splitting. 
With the high mobility and large Zeeman gap on the magnetic substrate CrGeTe$_3$, the MZMs at the corners of 1$T$-PtSe$_2$ monolayer are experimentally accessible.
Additionally, our two-BHZ model is proposed to capture the Majorana corner modes in the unconventional monolayer with anisotropy.
Our proposal for the Majorana corner modes in a 2D unconventional insulator/OAI can be widely studied in experiment.

\ \\
\noindent \textbf{Acknowledgments}
    This work was supported by the National Natural Science Foundation of China (Grants No. 11974395, No. 12188101), the Strategic Priority Research Program of Chinese Academy of Sciences (Grant No. XDB33000000), National Key R\&D Program of China (Grant No. 2022YFA1403800), and the Center for Materials Genome.

\appendix
\section{Band structures of PtSe$_2$ ribbons} \label{SM1}
\begin{figure*}[!htb]
\centering
\includegraphics[width=14 cm]{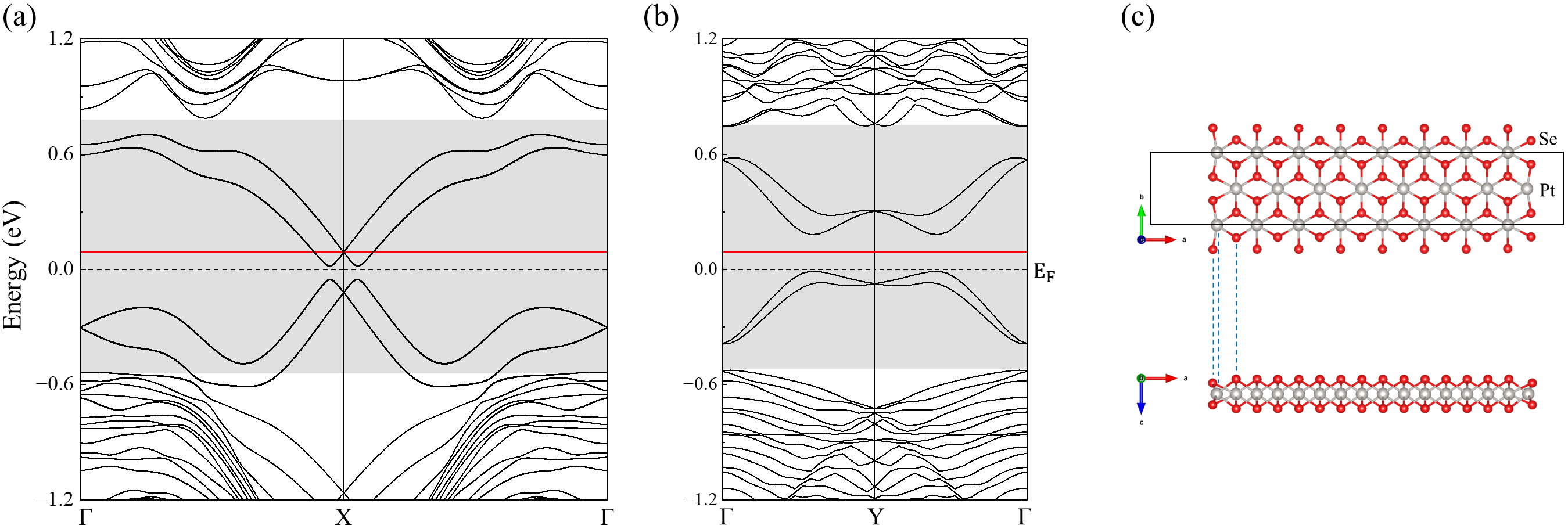}
\caption{(Color online)
Band structures with SOC of PtSe$_2$ ribbons with (a) zigzag edge and (b) armchair edge. 
(c) Top and side views of PtSe$_2$ ribbon with armchair edge. The black rectangle represents the unit cell. 
} \label{figS-zig+arm}
\end{figure*}

\begin{figure*}[!htb]
\centering
\includegraphics[width=17.5 cm]{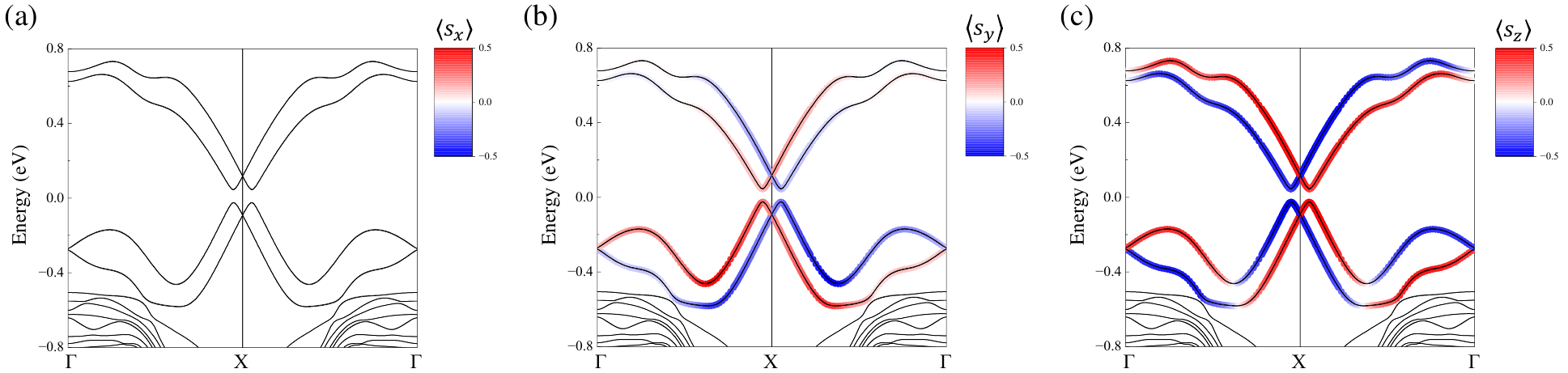}
\caption{(Color online)
Spin-resolved band structures of PtSe$_2$ ribbon with zigzag edge, considering SOC. The color scales represent the expectation values of spin components (a) $\ev{s_x}$, (b) $\ev{s_y}$, and (c) $\ev{s_z}$.
} \label{figS-spin}
\end{figure*}

The band structures with SOC of PtSe$_2$ ribbons with zigzag edge and armchair edge are shown in Fig.~\ref{figS-zig+arm}. The bulk band gaps (shadowed area) of the two ribbons are aligned in Figs.~\ref{figS-zig+arm}(a) and \ref{figS-zig+arm}(b). The degenerate zigzag edge states above the $E_F$ at the $X$ point are still located within the armchair edge gap (red line). 
In addition, the spin-resolved band structures of PtSe$_2$ ribbon with zigzag edge are presented in Fig.~\ref{figS-spin}, considering SOC. We can see that $\ev{s_x}=0$ and the $s_z$ component is dominant near the $E_F$.

\begin{figure*}[!htb]
\centering
\includegraphics[width=12 cm]{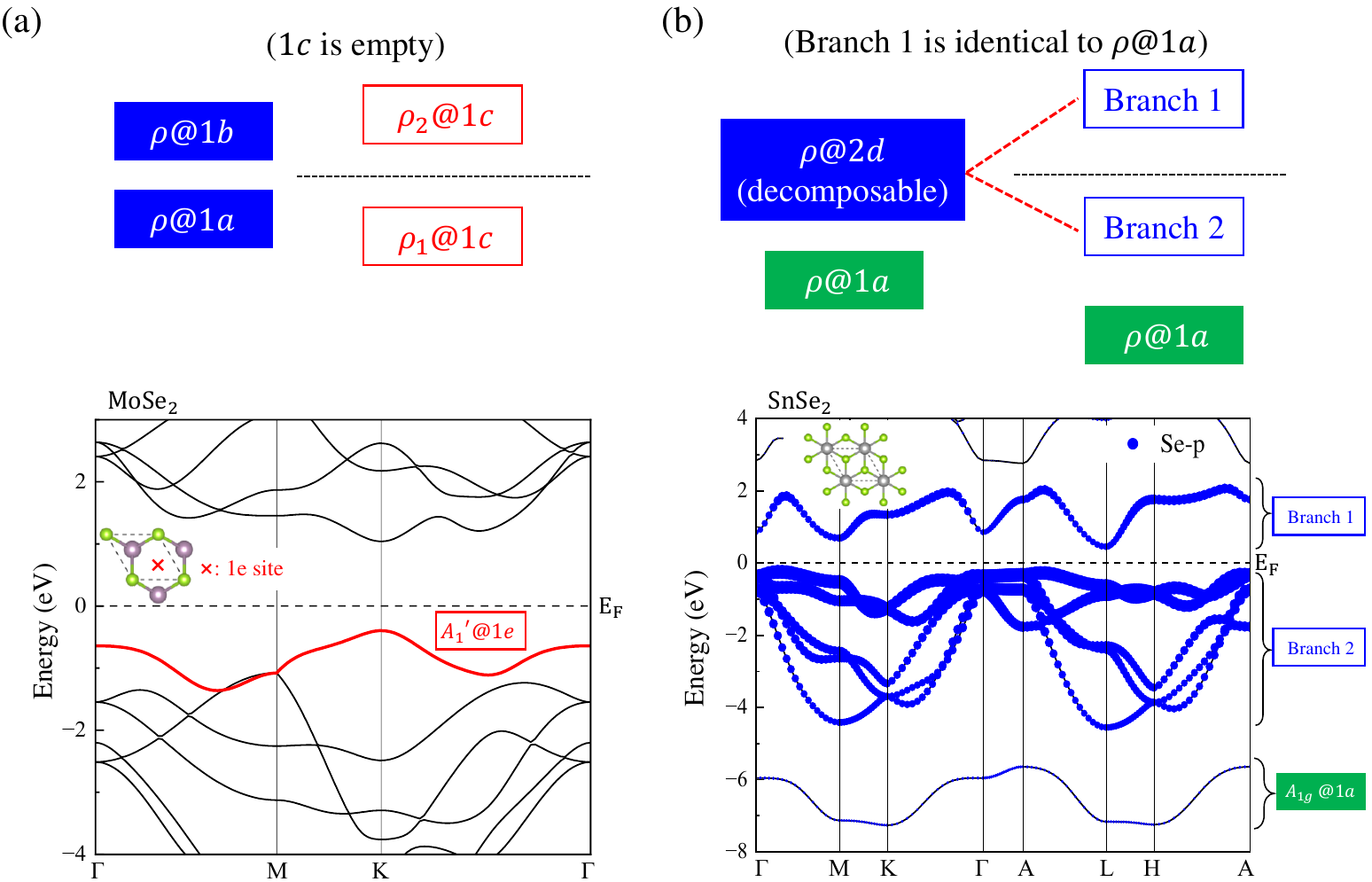}
\caption{(Color online) 
A schematic illustrating two types of the unconventionality of charge mismatch.
Case (a) with an empty-site EBR has the symmetry indicator, while case (b) cannot be diagnosed by BR analysis or symmetry indicator, termed SI-free unconventionality. 
The colors of the filled boxes indicate the atomic orbital characters, while the unfilled boxes represent the non-atomic states.
} \label{figS-type}
\end{figure*}

\begin{figure*}[!htb]
\centering
\includegraphics[width=16.5 cm]{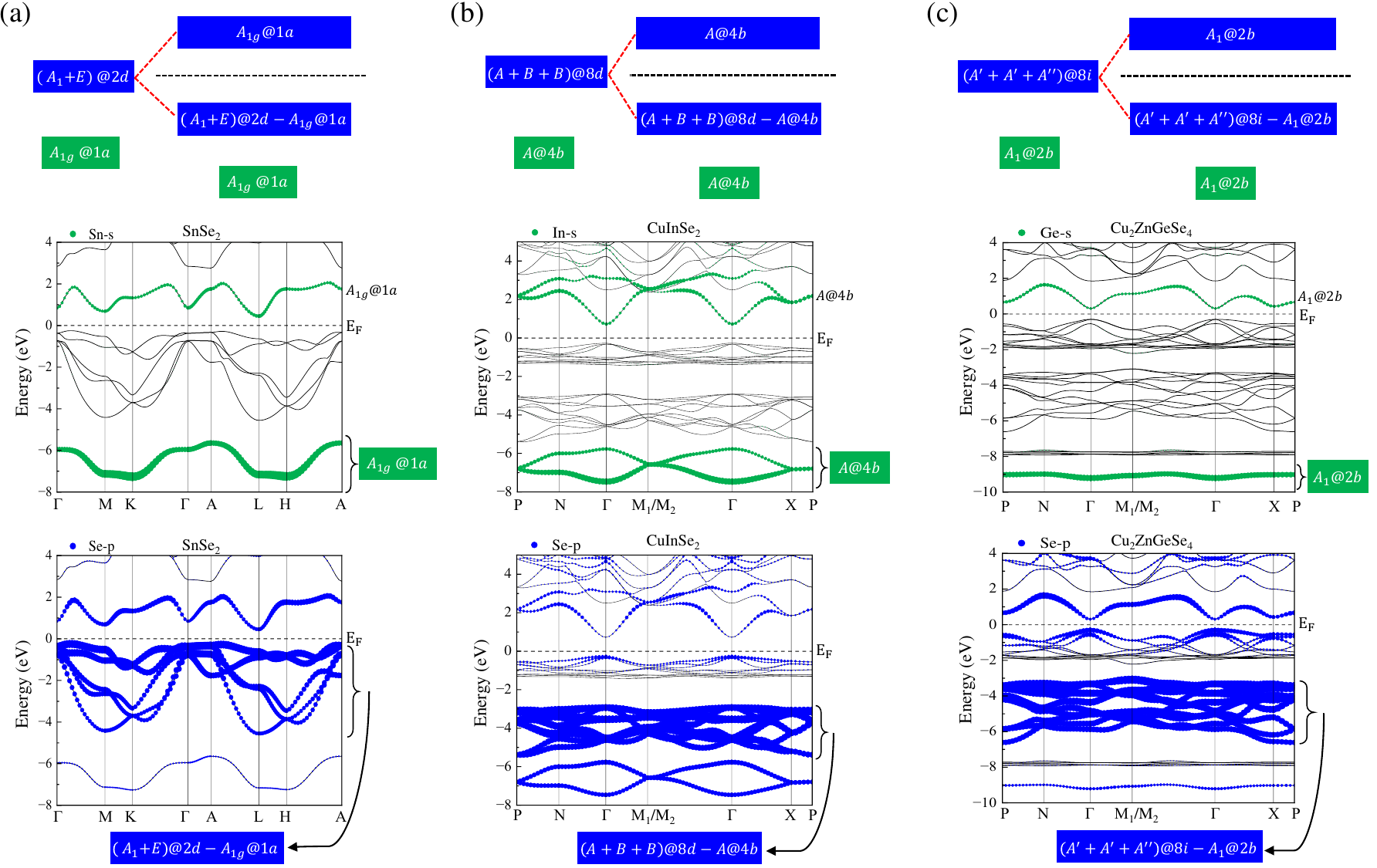}
\caption{(Color online)
The schematics of band hybridization and orbital-resolved modified Becke-Johnson (MBJ) band structures for (a,b) SnSe$_2$ (space group SG $P\overline{3}m1$), (c,d) CuInSe$_2$ (SG $I\overline{4}2d$), and (e,f) Cu$_2$ZnGeSe$_4$ (SG $I\overline{4}2m$). The sizes of green and blue circles represent the weights of corresponding $s$ and $p$ orbitals, respectively.
} \label{figS-type3band}
\end{figure*}

\section{More examples of SI-free unconventionality} \label{SM2}
At the atomic limit (without interatom hybridization), all the states originate from ABRs. After band hybridization in crystals, the electronic charge centers of occupied bands in an insulator can not be located at any atom, which is termed as unconventional insulator/OAI. Here we show two types of unconventionality: with the symmetry indicator [SI; Fig.~\ref{figS-type}(a)] and without SI [SI-free; Fig.~\ref{figS-type}(b)]. Case (a) with an empty-site elementary band representation (EBR, generators of the BRs) has the SI~\cite{aBR2022}, while case (b) cannot be diagnosed by BR analysis or symmetry indicator, termed SI-free unconventionality. The SI-free unconventionality is usually found in the compounds with (Ni/Pd/Pt)$^{4+}$, (Ga/In)$^{3+}$, or (Ge/Sn)$^{4+}$ valence states. 
Figure~\ref{figS-type3band} presents the schematics of band hybridization and orbital-resolved band structures for SnSe$_2$ (space group SG $P\overline{3}m1$), CuInSe$_2$ (SG $I\overline{4}2d$), and Cu$_2$ZnGeSe$_4$ (SG $I\overline{4}2m$), illustrating that their SI-free unconventional nature arises from the splitting of the decomposable BR with band hybridization.

\ \\
\ \\
%\bibliography{refs}

\end{document}